# Examining the assembly pathways and active microtubule mechanics underlying spindle self-organization

Lucan Yan[1], Tatsuya Fukuyama[1], Megumi Yamaoka[2], Yusuke T. Maeda[1,*], and Yuta Shimamoto[2,*]

[1] Department of Physics, Faculty of Science, Kyushu University, Motooka 744, Fukuoka 819-0395, Japan

[2] Physics and Cell Biology Laboratory, National Institute of Genetics & Department of Genetics, SOKENDAI University, Yata 1111, Mishima, Shizuoka 411-8540, Japan

*correspondence: ymaeda@phys.kyushu-u.ac.jp, yuta.shimamoto@nig.ac.jp

**ABSTRACT**

The bipolar organization of the microtubule-based mitotic spindle is essential for the faithful segregation of chromosomes in cell division. Despite our extensive knowledge of genes and proteins, the physical mechanism of how the ensemble of microtubules can assemble into a proper bipolar shape remains elusive. Here, we study the pathways of spindle self-organization using cell-free Xenopus egg extracts and computer-based automated shape analysis. Our microscopy assay allows us to simultaneously record the growth of hundreds of spindles in the bulk cytoplasm and systematically analyze the shape of each structure over the course of self-organization. We find that spindles that are maturing into a bipolar shape take a route that is distinct from those ending up with faulty structures, such as those of a tripolar shape. Moreover, matured structures are highly stable with little occasions of transformation between different shape phenotypes. Visualizing the movement of microtubules further reveals a fraction of microtubules that assemble between extra poles and push the poles apart, suggesting the presence of active extensile force that prevents pole coalescence. Together, we propose that a proper control over the magnitude and location of the extensile, pole-pushing force is crucial for establishing spindle bipolarity while preventing multipolarity.

**INTRODUCTION**

Cell division is a highly dynamic process during which replicated genomes are spatially segregated into the newly-created two daughter cells. The segregation is driven by the spindle, a micron-sized force-generating machinery assembled from microtubule cytoskeletal polymers (1). The bipolarity of the spindle is essential for the transport of each replicated genome toward two opposite ends of the dividing cell. Failures in the process of spindle assembly are linked to diseases, such as cancer, and several developmental disorders, such as Down's syndrome (2, 3).

Spindle assembly is generally considered as the process of self-organization, as it requires no template or blueprint that defines the structure's overall morphology, but rather the ordered large-scale architecture can autonomously emerge from interactions of smaller elements. This is well documented by a seminal work done by Heald and colleagues, who have shown that spindles of bipolar shape can be formed around DNA-coated microbeads without major structural cues, such as pairs of centrosomes and pairs of kinetochores (4). Rather, the spindle organization can be promoted by actions of force-generating motor proteins, which locally crosslink and slide microtubules while the large-scale architecture is assembled (5). Past studies have explored the mechanisms of spindle self-organization using genetic and biochemical approaches, by perturbing molecules and interactions of molecules and analyzing their impacts on spindle morphologies (6). Advanced live-cell imaging assay further allowed for visualizing temporal dynamics of the microtubule self-organization process and the emergence of spindle bipolarity (7-9). However, we still have poor quantitative knowledge on the pathways and fates of microtubule structures that dynamically assemble, grow, and change their shapes throughout the course of spindle self-organization. Moreover, as cells typically have considerable biochemical heterogeneity (10), it has been a challenge to link the observed differences in spindle assembly dynamics with their physiological causes in a molecularly unperturbed setting. Hence, in contrast to our extensive knowledge of molecules and their link to phenotypes, the way in which microtubules self-organize into the defined bipolar morphology remains enigmatic.

An important feature that distinguishes the spindle from other self-organization systems is that the major elements of the structure (i.e., microtubules) are highly dynamic due to their rapid turnover. The nucleation, polymerization, and depolymerization can take place on the order of tens of seconds (11) whereas the spindle assembly requires approximately an hour to complete. Besides, individual microtubules undergo an active directional flow toward assembled poles, called the poleward flux (12), whose speed (~3 µm/min) is much faster than the growth rate of the spindle structure (<0.5 µm/min). Previous *in vitro* studies have reported that a mixture of microtubules and microtubule-based motor proteins can generate large-scale ordered structures as a result of collective polymer motility (13-15). The patterns generated include asters and nematic arrays (16-19), which are recurring motifs of the spindle (poles and the equator, respectively). These studies have demonstrated that cytoskeletal self-organization can be driven by integrated local interactions of polymers without global control over microtubule activities and helped develop physical models to account for the emergence of such patterns. However, the assays were mostly done with static polymers such as Taxol-stabilized microtubules. Hence, our understanding of spindle self-organization is still far from reach particularly regarding how the matured spindle structure can be built and maintained over a long timescale while microtubules move and turnover with much shorter timescales.

Here, we investigate the dynamics of spindle self-organization using Xenopus egg extracts, a powerful cell-free system that can recapitulate diverse cell cycle events including nuclear and cytoskeletal assembly. Hundreds of spindles could be assembled in a bulk cytoplasm with which the morphological dynamics of growing structures could be simultaneously captured using large-field fluorescence imaging. By applying a quantitative analysis method that could extract the morphological features of the assembly "intermediates," which can be observed during the spindle self-organization process, we reveal two distinct pathways that each lead to bipolar or tripolar shape phenotype. Visualizing the internal movement of microtubules further reveals a subset of antiparallel polymer bundles that lie between pairs of poles in single tripolar spindles, suggesting active cytoskeletal mechanics that push the poles apart and prevent the pole coalescence. Together, we discuss how the control over local microtubule assembly ensures the proper spindle self-organization, which is required for successful chromosome segregation in cell division.

## RESULTS

**Establishing a method for analyzing the pathways of spindle self-organization**

We started the spindle self-organization reaction by adding demembranated sperm DNA to freshly-prepared Xenopus egg extracts such that thousands of nuclei could be assembled in the bulk interphase cytoplasm. The extract was then cycled back into a mitotic phase to trigger the disassembly of the nuclei, and then immediately enclosed in an imaging chamber such that the subsequent microtubule organization process could be visualized using confocal fluorescence microscopy (Fig. 1A). As shown in the time-lapse images (Figs. 1B and 1C), after the nuclear envelope break-down ($t = 0$), microtubules first organized into a round-shaped small aggregate ($t = 14$ min) and then grew to an anisotropic structure with partially focused poles ($t = 28 – 56$ min); finally, the structure reached a steady-state at which either a bipolar (Fig. 1B) or non-bipolar shape, such as one with three poles (Fig. 1C), were visually distinguishable ($t = 70$ min). These assembled structures coexisted in the same bulk cytoplasm and maintained a steady-state for tens of minutes (i.e., compare spindle shapes at $t = 56$ min and 70 min in each time-lapse). Although the matured spindle structures often exhibited transient local deformation (e.g., bending), their overall shape phenotypes were stably maintained in most assembly cases (N = 207 out of 237). For example, a spindle that had been assembled into a tripolar shape dynamically changed the relative position of its poles but it did not transform into bipolar (Movie S1). The overall assembly process typically took 45-60 min with ~70% of the total structures to be bipolar, being consistent with the time and fraction that can be achieved with our typical biochemical assays performed in test tubes (20).

To quantitatively analyze how microtubules are organized into the defined morphologies (e.g., bipolar, tripolar), we analyzed the shape of the "assembly intermediates," which were captured in the middle of the self-organization process (e.g., at 14, 28, and 42 min of the time-lapse images in Fig. 1). For this purpose, we employed multi-pole expansion analysis (21, 22), which quantitatively extracts the morphological features of an object by dissecting their outlines into a series of discrete shape modes that are comprised of orthogonal sinusoidal functions (Fig. 2A). The shape of the microtubule-based structures, which could be observed at each time point in the time-lapse sequence (interval: 1 min), was analyzed by tracing the outline with a vector from the centroid of the spindle (defined as vector $\boldsymbol{R}$, Fig. 2A). The amplitude of each shape mode, $C_n(t)$, is defined as:

$$C_n(t) = \frac{1}{2\pi} \int_{-\pi}^{\pi} R(\theta, t) e^{-in\theta} d\theta \quad [\text{Eq. 1}]$$

$$(n = 0, 1, 2, 3 \ldots \quad -\pi \leq \theta \leq \pi)$$

where $R(\theta, t)$ represents the distance from the spindle's centroid to its boundary at angle $\theta$ and time $t$. The absolute values of $C_0$, $C_2$, and $C_3$ are quantitative indices that correspond to monopolar/circular, bipolar, and tripolar shapes, respectively. Here, $C_1$ is negligible because it corresponds to the distance between the spindle's centroid and the origin of the vector $\boldsymbol{R}$, which are identical in our analysis. The amplitude of each shape mode can be represented by $R_n(\theta) = C_n e^{in\theta}$ (Fig. 2A). For the proof of concept, we analyzed three representative spindle images, bipolar, tripolar, and monopolar/circular, using this method (Figs. 2B, 2C, and 2D). The bipolar spindle yielded the larger $C_2$ value (for the amplitude of $R_2(\theta)$) than the $C_3$ value (for the amplitude of $R_3(\theta)$) whereas the tripolar spindle yielded the larger $C_3$ value than the $C_2$ value. We then calculated the normalized shape parameters $\tilde{C}_2 = C_2/C_0$ and $\tilde{C}_3 = C_3/C_0$, whose magnitudes respectively indicate the tendency of bipolar and tripolar shapes that arise in each assembled structure independent of the structure's size.

**Bipolar and tripolar spindle self-organization occurs through distinct assembly pathways**

We collected a series of time-lapse sequence data for a total of N = 115 spindles and performed the above-mentioned multi-pole expansion analysis. We found that trajectories constructed by $\tilde{C}_2$ and $\tilde{C}_3$ exhibited several distinct patterns in the two-dimensional plane, such as one showing a strong upward movement (along the $\tilde{C}_2$ axis) and one drifting towards the right (along $\tilde{C}_3$ the axis) with a large fluctuation (Fig. S1). We thus analyzed how each of these distinct trajectories reaches the matured, steady-state spindle shape, by objectively classifying the final spindle morphology using a pre-trained Linear Discriminant Analysis (LDA) model. LDA model was developed based on a machine-learning algorithm (see Materials and Methods) (23, 24). We first trained the algorithm by using a total of 200 representative spindle images showing distinct morphologies. These images were subjected to visual inspection and manually classified into three cases: circular, bipolar, or tripolar. LDA model was then used to create a phase map overlaid onto the $\tilde{C}_2$ and $\tilde{C}_3$ plane, with three distinct regions that each corresponds to the degree of multipolarity (background gray areas, Fig. 2E). This phase map allowed us to sort the set of time-lapse data into three distinct groups, i.e., circular, bipolar, and tripolar (colored plots, Fig. 2E). Finally, each time-lapse sequence sorted in this way was used for characterizing the spindle self-organization pathways. We note that our analysis yields the similar low $\tilde{C}_2$ and $\tilde{C}_3$ values for two types of structures, one with a circular shape, which we typically observed at an early phase of self-organization (e.g., at 14 min in Figs. 1B and 1C), and one with monopolar (25), which appeared as a steady-state structure with single focused poles at much later time points (e.g., Fig. 2D). We thus label the bottom left region in the $\tilde{C}_2$–$\tilde{C}_3$ plane "monopolar/circular" (white area, Fig. 2E).

On establishing the imaging assay and the shape analysis, we examined how the matured, steady-state spindle shape phenotypes are linked to the self-organization pathways. We found that the trajectories of the structure's growth, as depicted in the $\tilde{C}_2$–$\tilde{C}_3$ plane, are distinctly different among samples that reached either bipolar, tripolar, or monopolar/circular (Fig. 3). Specifically, many samples with the final spindle morphology of bipolar shape exhibited a strong upward movement with a slight deviation toward right, followed by an up-left movement until $\tilde{C}_2$ prevails over $\tilde{C}_3$ (N = 90 out of 115) (Fig. 3A). On the other hand, in the case of tripolar, the trajectories highly fluctuated with significant positive slopes particularly at an early stage of the self-organization, before reaching the region where $\tilde{C}_3$ is predominant (N = 20 out of 115) (Fig. 3B). The rest fraction, which was classified as monopolar/circular (N = 5 out of 115), also exhibited large fluctuation but was converged to the region with small $\tilde{C}_2$ and $\tilde{C}_3$ (Fig. 3C). The positive slope that appears in the $\tilde{C}_2$–$\tilde{C}_3$ plane indicates that both bipolar and tripolar modes increase concomitantly. On the other hand, the negative slope in the plane indicates that bipolar and tripolar are mutually exclusive; i.e., the appearance of bipolar feature suppresses the tripolar tendency and *vice versa*. For example, in the bipolar case (Fig. 3A), the structure first exhibits the growth of both bipolar and tripolar features and then switches to a predominant bipolar mode as the assembly progresses.

**Mutual inhibition between bipolar and tripolar growth modes results in the formation of steady-state bipolar phenotype**

The distinct self-organization pathways as revealed above motivated us to examine whether any fundamental relationships exist between the dynamics of $\tilde{C}_2$ and $\tilde{C}_3$. To this end, we calculated cross-correlation between $\tilde{C}_2$ and $\tilde{C}_3$, which indicates the time-dependent correlation between bipolar growth and excess pole formation. Here, the cross-correlation function $Cr(\Delta t)$ is defined as

$$Cr(\Delta t) = \frac{\langle \tilde{C}_2(t) - \langle \tilde{C}_2 \rangle \rangle \cdot \langle \tilde{C}_3(t + \Delta t) - \langle \tilde{C}_3 \rangle \rangle}{\sqrt{\langle (\tilde{C}_2(0) - \langle \tilde{C}_2 \rangle)^2 \rangle \langle (\tilde{C}_3(0) - \langle \tilde{C}_3 \rangle)^2 \rangle}} \qquad [\text{Eq. 2}]$$

, where $\Delta t$ represents the time-delay of tripolar mode ($\tilde{C}_3$) relative to the bipolar mode ($\tilde{C}_2$). We found that for bipolar spindles, this cross-correlation function yielded predominant negative values at any lag-time $\Delta t$ (Fig. 3D). This is mainly due to the characteristic trajectory of bipolar spindles that directs to the bipolar region while suppressing tripolar tendency (i.e., $\tilde{C}_2$ becomes larger while $\tilde{C}_3$ becomes smaller in Fig. 3A). Furthermore, the peak of the correlation $Cr(\Delta t)$ appeared at $\Delta t = 0$ min (i.e., no time-delay), indicating that the growth of bipolar shape occurred simultaneously with the suppression of tripolar shape (Fig. 3D). On the other hand, a prominent positive correlation was found in tripolar spindles (Fig. 3E). This is because the trajectory of tripolar spindles in the $\tilde{C}_2$–$\tilde{C}_3$ plane directs to the tripolar region where both $\tilde{C}_2$ and $\tilde{C}_3$ are large (Fig. 3B). Importantly, we find a significant time delay in the cross-correlation in the tripolar case (i.e., the peak is shifted by ~13 min toward $\Delta t > 0$, Fig. 3E), suggesting that a growth of bipolar shape proceeds the growth of tripolar shape. Finally, as for the monopolar/circular case, the correlation $Cr(\Delta t)$ was less profound at any lag-time $\Delta t$ compared to bipolar and tripolar cases (Fig. 3F). Together, our cross-correlation analysis reveals the distinct morphological growth dynamics that are characteristic to each of the matured spindle phenotypes. More specifically, the bipolar phenotype can be achieved through the mutual competition of bipolar and tripolar assembly forces. On the other hand, the tripolar phenotype emerges not due to the loss of bipolar versus tripolar competition, but rather associate with a delayed growth of an excess pole. The monopolar/circular phenotypes can be achieved by an uncoordinated activity of bipolar and tripolar assembly forces.

**Bidirectional microtubule flux that appears between poles may stabilize tripolar spindle morphology**

To understand the underlying microtubule mechanics, we performed fluorescence speckle microscopy, a method that can visualize the movement of individual polymers and their coordinated "flow" in cytoskeletal structures (Fig. 4) (26). The method relies on a small amount of dye-labeled tubulins, which were added to extracts such that these tubulin subunits could be stochastically incorporated into the lattices of microtubules and form "speckles," serving as fiduciary markers of the polymer movement. The movement of individual speckles was analyzed using particle tracking velocimetry we performed previously (27).

Tubulin speckles moved highly stochastically while the assembled structure underwent a slow translational drift across the cytoplasm, making it difficult to determine the absolute movement of individual speckles in the structure. We therefore divided the spindle self-organization process into six successive phases by referencing the shapes of the assembly intermediates that were mapped in the $\tilde{C}_2$–$\tilde{C}_3$ plane (Figs. 4A-4C). The analysis was performed both for bipolar and tripolar assembly cases. At Phase 1 of both cases, the trajectory starts from the region in which the structure's shape is classified as monopolar/circular (blue plots, Figs. 4B and 4C). At Phases 2 and 3, the trajectory crosses the border between circular and tripolar (orange and yellow plots, Fig. 4B, C). When entering Phase 4, the value of $\tilde{C}_2$ drastically increases in the bipolar case and the structure's shape becomes bipolar (purple, Fig. 4B). On the other hand, in the tripolar case, the trajectory highly fluctuates and the value of $\tilde{C}_3$ is kept high (purple, Fig. 4C). Finally, at Phases 5 to 6, the structures reach a steady state. The overall spindle shape becomes distinguishable after passing Phase 4.

We hypothesized that there exists a difference in the movement of microtubules at the early self-organization phase (e.g., before Phase 4), in a way that such a difference could determine the fate of spindle morphologies. To test this, we analyzed the orientation of speckle movement at the early phases before a noticeable morphological feature was recognizable (Figs. 4D-Q). The orientation $\phi$ was defined as the angle relative to the major axis of microtubule flux, which was determined based on the peak of the angle distribution for all speckles analyzed (black arrows, Figs. 4D and 4K). The speckles analyzed were ones located around the structure's center (colored regions, Figs. 4D and 4K), at which many microtubules nucleate, polymerize, and move towards varying directions. Unexpectedly, we found no significant difference in the microtubule movement between bipolar and tripolar assembly cases at Phases 1-3 (green and red bars, Figs. 4E-G and L-N). A prominent difference was found only after Phase 4, at which the feature of bipolarity also becomes prominent (Figs. 4H and 4O). At this stage, the speckles of the bipolar assembly exhibited the flux toward either of the two poles with a highly polarized motility at the orientations of $\phi = 0$ and $\pi$ (Figs. 4I and 4J). On the other hand, for the tripolar assembly case, the tendency of such a polarized flux was less profound (Figs. 4P and 4Q). This implies that the local alignment of microtubules around the structure's center, which was expected to occur at the early phases of self-organization, is not the predominant mechanism determining the final spindle morphology. The mechanism that can stabilize the assembly phenotypes may thus exist in other spindle location.

Hence, we sought to analyze the speckle dynamics away from the structure's center. Our primary focus was on the region between focused poles because prevention of pole coalescence could result in the stabilization of multipolar phenotypes. We found that a tripolar spindle, which had reached a steady state and maintained its overall morphology over tens of minutes, exhibited a transient rotational motion of one pole relative to the other two poles (Figs. 4R and 4S). We analyzed how internal microtubule mechanics drive this movement, by quantifying the direction and velocity of individual speckles near the excess pole (Fig. 4T). We found that, in addition to the overall persistent flow of speckles directing from the structure's center toward each pole (typical speed is 2-4 µm/min), there exists a strong bidirectional flow that moves at a speed of 10-12 µm/min between two of the three poles (top and right poles) (highlighted gray background, Fig. 4T). The direction of this predominant speckle flow is consistent with the direction of one pole moving away from the other, suggesting the assembly of microtubule arrays which generate a pushing force to prevent the poles from coalescing. An additional time-lapse data of a different tripolar spindle (Fig. S2) shows a similar microtubule flow pattern between poles, which supports the presence of bidirectional flow across poles.

**DISCUSSION**

Using a combination of cell-free extracts, wide-field live fluorescence imaging, and a computer-based quantitative shape analysis, we found that spindle self-organization can be achieved through distinct assembly pathways that each lead to different architectural morphologies such as bipolar and tripolar shapes, as summarized in Fig. 5A. The bipolar organization appeared to require a predominant growth of bipolar mode that simultaneously suppresses tripolar mode (upper path, Fig. 5A). On the other hand, the tripolar organization was found to occur associated with the promotion of excess pole formation that becomes prominent minutes after the growth of bipolar mode (lower path, Fig. 5A). We also found that the matured structures are highly stable and can rarely transform into another shape phenotypes (double arrows, Fig. 5A).

Our analysis, revealing the speckle flow of microtubules that persistently occurs across bipolar and tripolar spindles (Fig. 4), proposes a model for how the multipolarity is stabilized by active microtubule cytoskeleton (Fig. 5B). It has been shown that Xenopus extract spindles are comprised of many short microtubules (28) and thus the long-range order of polymer alignment should be achieved mainly through local interaction of overlapping polymers, which are mediated by microtubule-crosslinking proteins. Our observation reveals that a fraction of microtubules can be assembled between poles of a tripolar spindle and move predominantly along the axis connecting the two poles (Fig. 4T). This type of microtubule arrays is most likely antiparallel as it exhibits a bidirectional flow and thus should generate an active extensile force between the poles. The extensile force can be derived from microtubule-based motor proteins such as kinesin-5, which can crosslink antiparallel microtubules and push them apart with their minus-end leading while walking towards the polymers' plus-ends (inset, Fig. 5B) (29). Further, the magnitude of the pushing force can scale with the extent of the antiparallel microtubule overlap that kinesin-5 forms by crosslinking (30), thus exerting a larger repulsive force as the poles move closer to each other and the overlap length increases accordingly. It has been shown that the network of microtubules in the extracts are intrinsically contractile, and thus the poles can coalesce (31). This contractile force, which can be exerted by another class of motor proteins such as dynein (32-34) and Ncd (35), would be opposed by the extensile force activity we propose here, preventing the pole coalescence and stabilizing the faulty tripolar phenotype. Although this interpolar microtubule fraction could be observed at the steady state but not much in the assembly intermediates, most likely due to its relatively weak signal, such antiparallel bundles may also be formed at early phases of spindle self-organization and exert an active extensile force, preventing the transformation of tripolar structures into bipolar. Importantly, an extensile force is also required for the establishment and maintenance of spindle bipolarity by pushing apart two poles (36). We propose that proper control over the magnitude of this extensile force is crucial for assembling bipolar spindles as well as suppressing multipolarity.

The fraction of bipolar spindles assembled over other spindle phenotypes was >75% (Fig. 2E), indicating that the activity of the extract cytoplasm prefers to form bipolar structures. If components of the spindle turnover rapidly and the cytoplasm has a strong tendency to organize microtubules into a bipolar shape, the feature of the multipolarity, which is mistakenly generated at an early assembly phase, should cease as time progresses. Yet, the pathway to multipolar phenotypes (e.g., tripolar) was found to be highly stable. We propose that continuous nucleation and polymerization of new microtubules can retain the architectural information of the pre-assembled structure, which had been built from old microtubules polymerized at earlier time points, and lead to the further growth of the structure while maintaining its morphological feature. Once a multipolar structure is formed, newly assembled microtubules use the structure as a template and align along the pre-existing microtubules; the orientation of new microtubules is thus "locked" in the corresponding direction. This makes it difficult for the assembled structure to "escape" from the faulty shape pattern and to "correct" the morphology to bipolar. An important experiment to test this prediction would be to mechanically perturb the pre-existing structure (i.e., template) and analyze whether the structure's growth direction can be altered. This could be achieved by biophysical micromanipulation techniques, such as those based on force-calibrated microneedles (37, 38) and atomic force microscopy (39). Elucidating the active cytoskeletal mechanics underlying spindle assembly should help advance our understanding of cell division and provide insights into the general principle of self-organization in biological systems.

## MATERIALS AND METHODS

**Preparation of cytoplasmic extracts.**

Cell-free extracts we employed for our spindle self-organization assays were prepared from unfertilized eggs of *Xenopus laevis*, as essentially described in (40). For each preparation, 2-3 female frogs were primed using progesterone (hor-272, Prospec) and then induced to ovulate using human chorionic gonadotropin (CG-10, Sigma). After ~16 h from the gonadotropin injection, eggs were collected and washed with MMR (5 mM Na-Hepes, 0.1 mM EDTA, 100 mM NaCl, 2 mM KCl, 1 mM $MgCl_2$, and 2 mM $CaCl_2$, pH 7.7), treated with 2% cysteine solution, and then rinsed with buffers in the following order: XB (10 mM K-Hepes, 100 mM KCl, 1 mM $MgCl_2$, 0.1 mM $CaCl_2$, and 50 mM sucrose; pH 7.7), CSF-XB (XB plus 1 mM $MgCl_2$ and 5 mM EGTA), and CSF-XB + PI (CSF-XB plus 10 μg/ml each of leupeptin, pepstatin A, and chymostatin). The rinsed eggs were supplemented with 10 μg/ml cytochalasin D, packed in centrifuge tubes (344057, Beckman) using a tabletop centrifuge (5702R, Eppendorf), and then mechanically lysed using a SW-55 rotor at 10,000 × g for 15 min at 16°C (Optima XE-90, Beckman). Following centrifugation, a cytoplasmic fraction was retrieved from the tubes using a 16-G needle and supplemented with Energy mix (75 mM creatine phosphate, 1 mM ATP, and 1 mM $MgCl_2$). The protease inhibitors and cytochalasin D were also added to the collected cytoplasm. The entire procedure was performed in a temperature-controlled room at 18 ± 1 °C. The extracts prepared were stored in 1.5-ml test tubes on ice and used within 6 hours.

**Spindle assembly and imaging.**

For spindle assembly, extracts prepared as above (30 μl each = 1 volume) were supplemented with demembranated sperm (at 1,000 nuclei/μl) and $CaCl_2$ (at 0.4 mM) in 1.5-ml test tubes and incubated at 18°C for 90 min to assemble interphase nuclei. Fresh CSF-arrested extract (1 volume; no sperm added) was then added to each reaction to cycle the extract back into the mitotic phase. Immediately after supplementing with X-rhodamine-labeled tubulin (10 μg/ml) (TL620M, Cytoskeleton, Inc.) and SYTOX green (S7020, Thermo), four μl of the mixture was squashed onto a glass slide with a 18 × 18 mm coverslip (Matsunami) and then sealed using VaLaP (Desai et al., 1999). Time-lapse imaging was performed using a 20× objective lens (1.20 NA, Plan Apo WI, Nikon), a spinning-disk confocal unit (CSU-X1, Yokogawa) with two excitation lasers (OBIS 488 and 561, Coherent), and a sCMOS camera (Neo4.1, Andor). Image acquisition was performed at 1-s intervals with the exposure of 500 ms whose timing was controlled using image acquisition software (NIS-Elements, ver. 4.5, Nikon). For assays with simultaneous tracking of many spindles assembled in the same extract reaction, a raster-scanning of a 5 × 6 area near the center of the sample chamber was performed using a large image mode, which was equipped in NIS-Elements software. For assays with speckle microscopy, the microscopy configuration as above was used but with a 60× objective lens (1.20NA, Nikon) instead of 20×, and single-area images were acquired for generating each time-lapse sequence. For speckle imaging assays, the labeled tubulin added was 200 ng/ml instead of 10 μg/ml.

**Image Processing and multipole expansion analysis.**

Before we analyzed the spindle shape by multipole expansion analysis, we employed deep learning algorithms to remove background noise using a toolbox of ImageJ (CSBDeep) (41). To detect the outline shape of spindles, a binarization processing was executed by a self-written MATLAB code (21). The binarization threshold was given by Otsu method (42) and the fold-change was applied to capture the appropriate spindle shape. After detecting the spindle area, the position of the center of mass was determined based on the binarized images. The distance from the center of mass to the outline of the spindle was then calculated for all angles. In this way, the shape of the spindle was analyzed at each time point of the time-lapse recording, $t$, and the spindle shape was expressed as $R(\theta, t)$ with the time $t$ and angle $\theta$ in Eq. [1] (see Main text). The amplitude of each multipole shape is described by $R_n(\theta) = C_n e^{in\theta}$ where the integer $n = 2$ or $n = 3$ corresponds to the number of poles and represents bipolar or tripolar. Higher modes ($n > 3$) are negligible because by using only 3 components ($n = 0$, 2, 3) are sufficient to characterize the shape of the spindle. The shape of spindles, therefore, can be expressed by the superposition of $R_0(\theta)$, $R_2(\theta)$ and $R_3(\theta)$, and the original shape of the spindle can be restored as a superposition of these amplitudes. To classify different spindle's phenotypes, we adopted a classification method based on Linear Discriminant Analysis (LDA) model, because of its simplicity and accuracy (23). We executed the LDA model classification analysis by using a custom-made MATLAB code (24). The model was first trained with two hundreds of randomly picked spindle images, comprised of 40 monopolar, 80 bipolar, and 80 tripolar samples.

**ACKNOWLEDGEMENTS**

This work was supported by Grant-in-Aid for Scientific Research on Innovative Areas JP18H05427 and Scientific Research (B) JP20H01872 (to Y.T.M.); Scientific Research (B) JP19H03201 and Challenging Research (Exploratory) JP20K21404 (to Y.S.). We are grateful for Prof. Akatsuki Kimura and his laboratory members for valuable inputs.

**CONFLICT OF INTEREST STATEMENT**

The authors declare no competing financial interests.

**AUTHOR CONTRIBUTIONS**

Y.S. conceived the study. M.Y. and Y.S. performed experiments. L.Y., T.F., and Y.T.M. performed data analysis. L.Y., Y.T.M, and Y.S. wrote the manuscript.

**FIGURES**

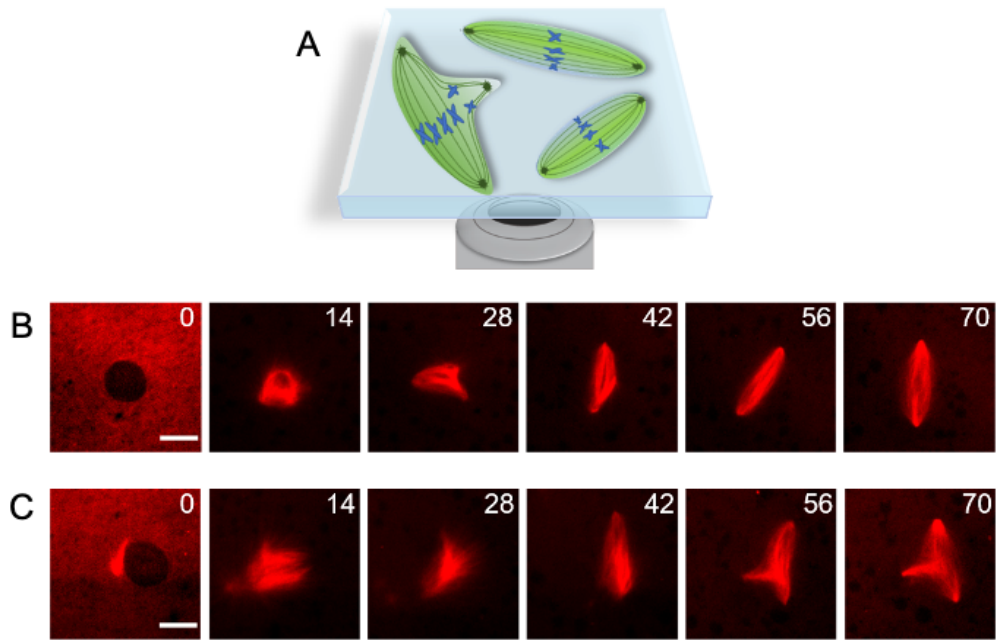

**Fig. 1 Examining spindle self-organization pathways. (A)** Schematic of the experimental setup. Spindles (green), which are assembled around replicated chromosomes (blue) in a bulk cytoplasm prepared from *Xenopus* unfertilized eggs, can be time-lapse imaged using confocal fluorescence microscopy in a glass-bottom experimental chamber. **(B, C)** Representative time-lapse images recording the process of spindle self-organization from the time at which nuclei were disassembled ($t = 0$ min) to the time several minutes after the structure reached a matured steady-state (t = 56 and 70 min) with bipolar (upper panels) **(B)** or tripolar shape (lower panels) **(C)**. Time stamp, min. Scale bars, 25 μm.

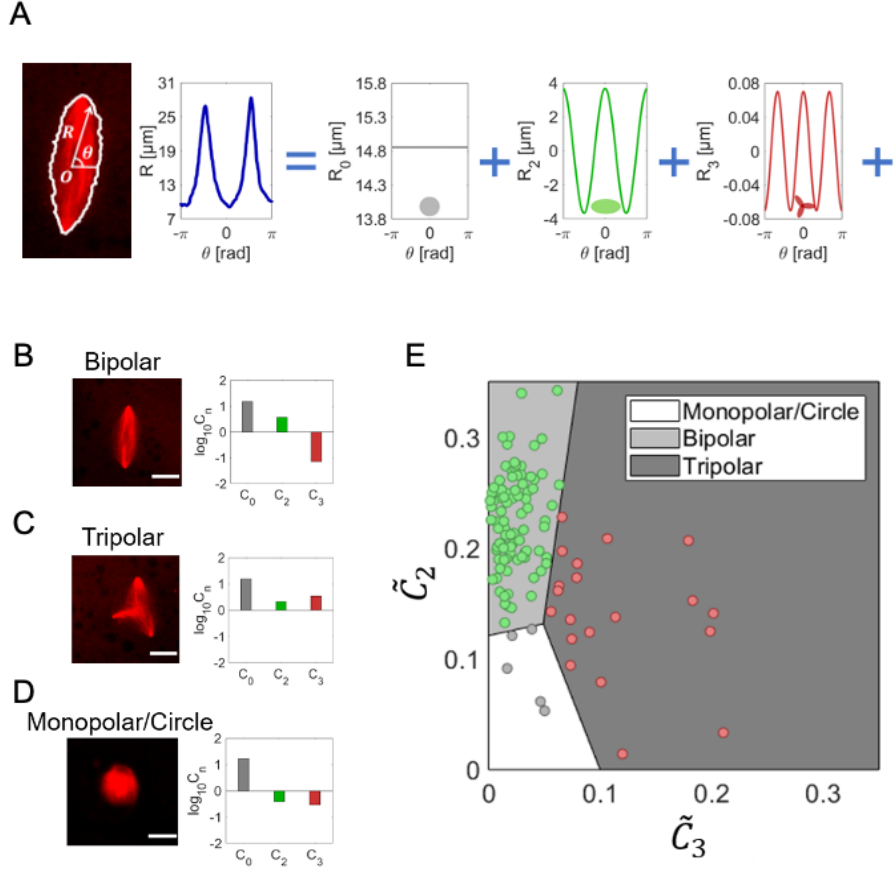

**Fig. 2  Shape analysis method and phase map defining spindle morphology. (A)** Representative fluorescence image of a spindle (red), in which the outline (white solid outline) was detected for shape determination. The point *O* indicates the centroid of the spindle calculated from the binarized image. The vector ***R*** represents the distance between the centroid and the spindle outline and the angle $\theta$ from the horizontal axis. $R_n(\theta) = C_n e^{in\theta}$ gives the magnitude of multipolar shapes (see Main Text): the 0-th component $C_0$ gives the average radius, the 2nd component $C_2$ gives the degree of bipolar shape, and the 3rd component $C_3$ gives the degree of tripolar shape. The spindle of a given shape can be decomposed into the discrete shape modes with $C_0$, $C_2$, and $C_3$. Higher-order modes were found to be negligibly small and thus omitted from subsequent analyses. **(B-D)** Representative fluorescence images of three different phenotypes, bipolar **(B)**, tripolar **(C)**, and monopolar/circular **(D)**, acquired at a steady-state. The panel next to each image shows the values of $C_0$, $C_2$, and $C_3$ calculated for each phenotype. **(E)** The phase map of $\tilde{C}_2$ and $\tilde{C}_3$. Supervised learning method was used to classify spindle phenotypes into three different groups: monopolar/circular (white area), bipolar (light gray area), and tripolar (dark gray area), which were obtained based on a pre-trained LDA model. Colored plots are the points at which each structure had reached at the end of self-organization process (*t* ~ 60 min) (N = 115). The analysis yielded 78.2% bipolar (green, N = 90), 17.4% tripolar (red, N = 20), and 4.4% monopolar (gray, N = 5).

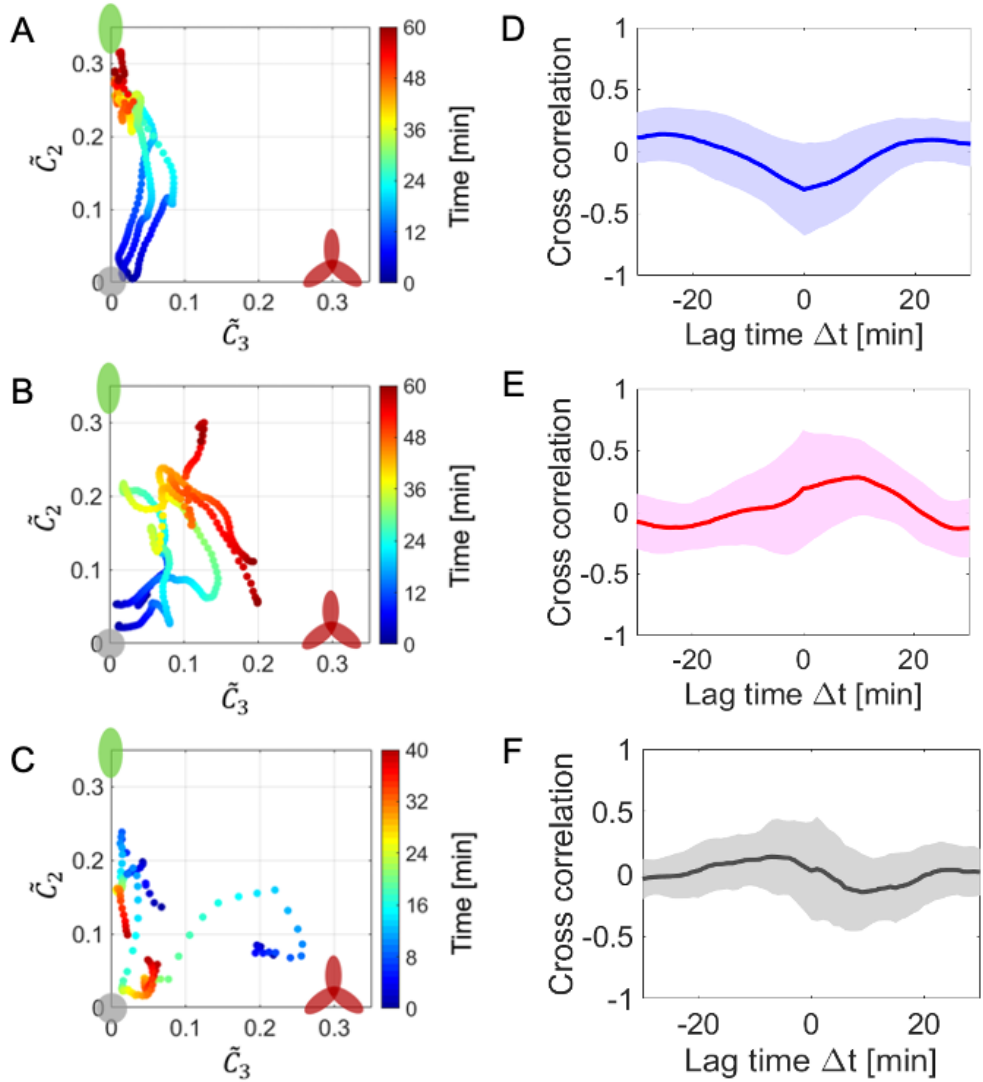

**Fig. 3 Distinct shape trajectories and time-dependent characteristics of spindle self-organization**. (A-C) Representative trajectories showing the time evolution of $\tilde{C}_2$ and $\tilde{C}_3$ for three bipolar assembly cases. Time-lapse images were used for the trajectory analysis. Marks of circle (gray), bipolar (green), and tripolar shape (red) are placed on near the graph's' corners to indicate spindle morphologies at certain area of the map. We set the origin of trajectory ($t$ = 0 min) to the time point at which the structure to be analyzed has formed a detectable surface. The trajectories of bipolar assembly cases **(B)**, tripolar assembly cases **(C)**, and monopolar/circular assembly cases **(D)** are shown (N = 3 examples each). **(D-F)** Cross-correlation between $\tilde{C}_2$ and $\tilde{C}_3$ was calculated for bipolar (an average of N = 77 samples with S.D.) **(D)**, tripolar (average of N = 32 samples with S.D.) **(E)**, and circular/monopolar samples (average of N = 19 samples with S.D.) **(F)**. The positive cross-correlation value indicates that both bipolar and tripolar modes grow simultaneously. The negative value on the other hand indicates that these two modes are mutually inhibited. The peak that appears with a positive lag-time indicates that the growth of bipolar mode proceeds the growth of tripolar mode.

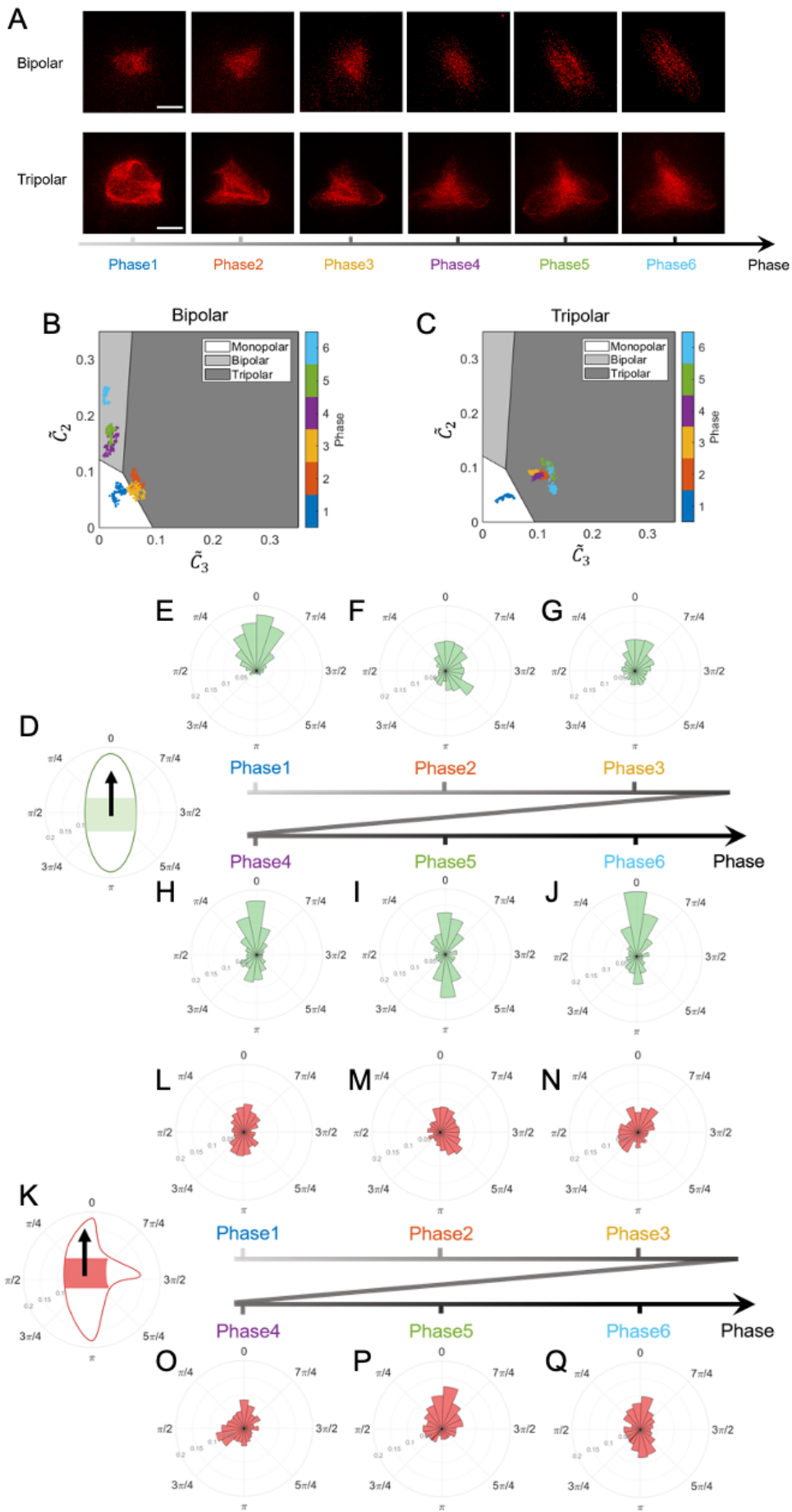

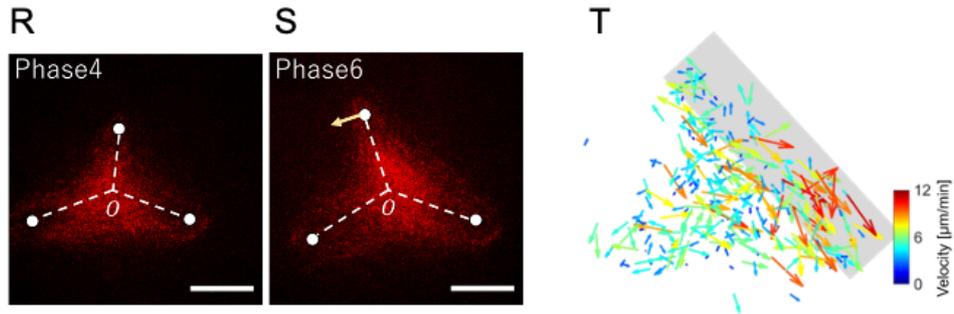

**Fig. 4 Analysis of microtubule flow using speckle microscopy during spindle self-organization. (A)** Representative fluorescence images of microtubule speckles in the time-lapse movies. Images in the upper panels are for a bipolar spindle and ones in the lower panels are for a tripolar spindle. Scale bars are 25 μm. The process of spindle self-organization was divided into six successive phases based on the structure's shape, which was mapped to the $\tilde{C}_2$-$\tilde{C}_3$ plane **(B, C)**. Colors indicate phases numbered from 1 to 6. **(D-N)** Speckle dynamics for the bipolar assembly case. Speckles at the structure's center (highlighted area in **D**) were analyzed for their movement orientation relative to the direction of predominant microtubule flow (black arrow in **D**). Circular histograms **(E-J)** show the distribution of speckle movement orientation at each self-organization phase. **(K-Q)** The identical analysis was performed for the tripolar assembly case. The schematic **(K)** and circular histograms for the orientation of speckle movement **(L-Q)** are shown. **(R, S)** Fluorescence images of speckles in a tripolar spindle (Phase 4 and 6). White dotted lines with circles indicate the axes and positions of poles. Yellow arrow indicates the movement of the top pole. Scale bars are 25 μm. **(T)** Vector map showing the orientation and velocity of speckles analyzed for the tripolar spindle. Warmer colors indicate faster speckle velocity. Gray rectangle highlights the area connecting the top and bottom right poles.

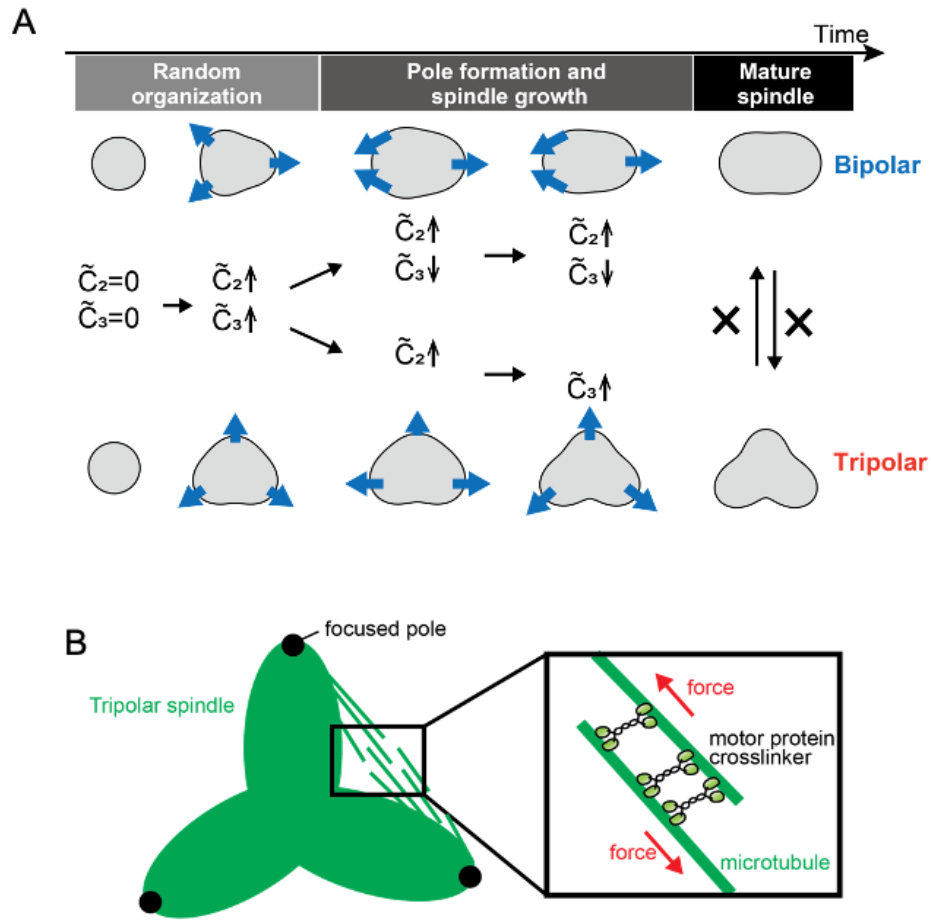

**Fig. 5. Model for spindle self-organization. (A)** Schematic illustration of the change in the spindle shape during self-organization. Upper panels are for bipolar assembly and lower panels are for tripolar assembly. The assembly pathways are distinct with different temporal growths of bipolar ($\tilde{C}_2$) and tripolar ($\tilde{C}_3$) modes. Blue thick arrows indicate the structure's predominant growth direction. Further, the matured, steady-state structures are stable and merely change their phenotypes (depicted as vertical arrows with crosses). **(B)** Proposed model of microtubule movement and associated force that stabilizes the tripolar assembly phenotype. The top and right poles tend to coalesce with each other due to the bulk cytoplasmic activity but are counteracted by opposing extensile force exerted within arrays of microtubules assembled between the poles (inset). This force can be exerted by plus-end directed, microtubule-crosslinking motor protein such as kinesin-5 (light green molecules in the inset).

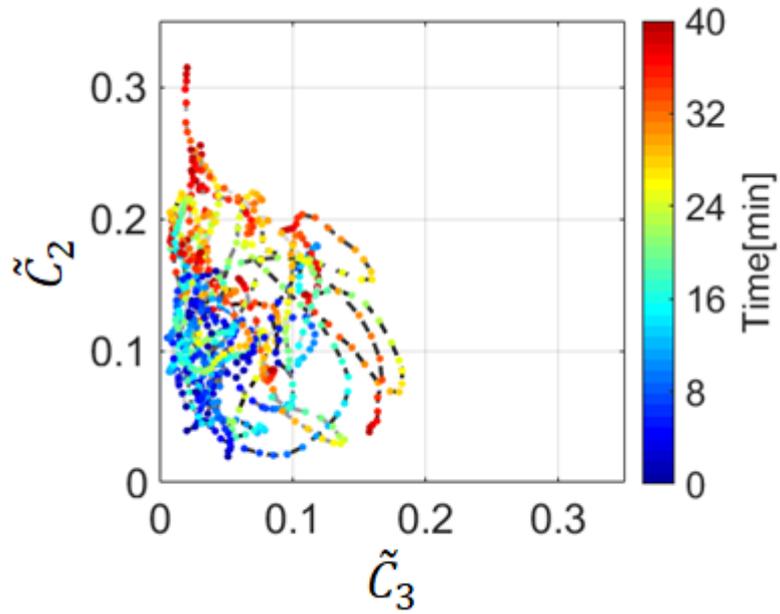

**Fig. S1. Trajectories of spindle self-organization pathways.** Trajectories of N = 20 samples were randomly chosen out of 115 and plotted in the $\tilde{C}_2$–$\tilde{C}_3$ plane.

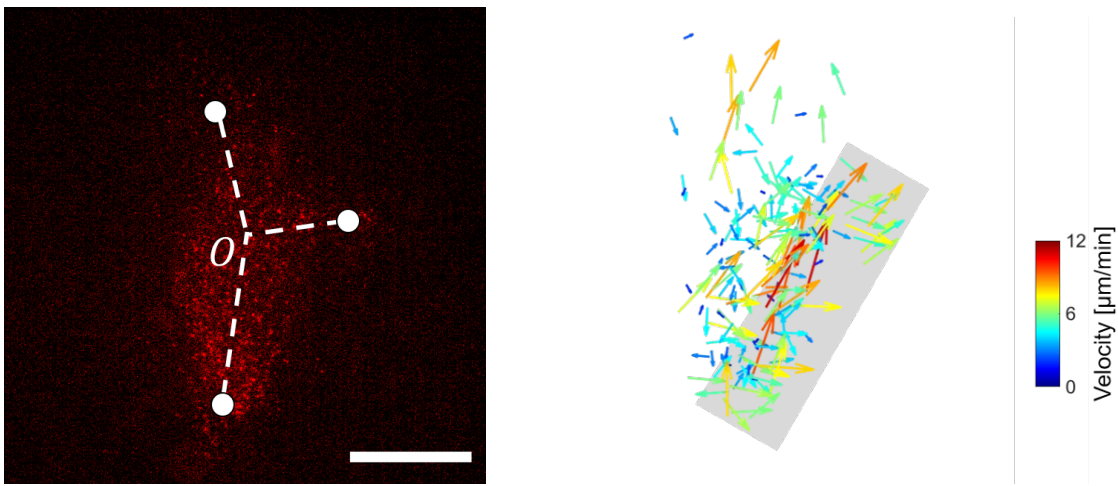

**Fig. S2. Additional data showing the strong bidirectional flow of microtubules between two poles of a single tripolar spindle.** The vector map was generated and presented as in Fig. 4T. Scale bar is 25 μm.